\documentclass[11pt]{article}
\usepackage[dvips]{graphicx}
\usepackage{enumitem}
\usepackage{amssymb}
\usepackage{color}
\usepackage{amsmath}
\usepackage{nicefrac}
\usepackage[left]{lineno}
\usepackage{hyperref}
\usepackage{adjustbox}
\usepackage{xspace}
\usepackage{multirow}
\usepackage{alphalph}
\usepackage[dvipsnames]{xcolor}
\usepackage{cite}

\definecolor{blu}{rgb}{0.,0.,1.}
\definecolor{red}{rgb}{1.,0.,0.}
\definecolor{burgundy}{rgb}{0.5, 0.0, 0.13}
\definecolor{crimsonred}{rgb}{0.6, 0.0, 0.0}
\definecolor{persianblue}{rgb}{0.11, 0.22, 0.73}
\definecolor{forestgreen}{rgb}{0.13,0.35,0.13}

\hypersetup{colorlinks, citecolor=crimsonred, linkcolor=persianblue, urlcolor=crimsonred}

\begin{document}
\centerline{\LARGE EUROPEAN ORGANIZATION FOR NUCLEAR RESEARCH}

\vspace{10mm} 

\vspace{-15mm}

\vspace{40mm}

\begin{center}
\boldmath
{\bf {\Large\boldmath{Search for heavy neutral leptons in $\pi^+$ decays to positrons }}}
\unboldmath
\end{center}
\vspace{4mm}
\begin{center}
\renewcommand{\thefootnote}{\fnsymbol{footnote}}
The NA62 Collaboration
\footnote[1]{
Corresponding authors: A.~Briano~Olvera, J.~Engelfried, \\
email: alejandro.briano.olvera@cern.ch, jurgen.engelfried@cern.ch
}
\end{center}

\begin{abstract}
A search for heavy neutral lepton ($N$) 
production in $\pi^+\to e^+ N$ in-flight decays using data collected by
the NA62 experiment at CERN in 2017--2024 is reported. Upper limits for the extended neutrino mixing
matrix element $|U_{e4}|^2$
are established at the level of $10^{-8}$ for heavy neutral
leptons with mass in the range 95--126\,$\mbox{MeV}/c^2$
and lifetime exceeding 50\,ns. 
\end{abstract}

\begin{center}
{\it accepted for publication in Physics Letters B}
\end{center}

\newpage

\section{Introduction}
\label{sec:intro}
While neutrinos are massless in the Standard Model (SM), the experimental evidence of neutrino oscillations~\cite{hirata1988experimental,ahmad2002direct} implies that neutrinos have a mass different from zero.
The neutrino Minimal Standard Model ($\nu$MSM)~\cite{asaka2005numsm,asaka-2} is a SM extension which can simultaneously explain neutrino oscillations, dark matter and the baryon asymmetry of the universe by adding three right-handed neutrinos, also called heavy neutral leptons (HNLs), which generate the non-zero masses of the active neutrinos via the see-saw mechanism~\cite{MINKOWSKI1977421}. The HNL masses are predicted to be below the electroweak scale, with the lightest HNL being a dark matter candidate in the keV/$c^2$ mass range~\cite{asaka-2}. 
The lifetime of the HNLs is expected to be long enough that they can be considered stable in production-search experiments.

HNLs with masses below $(m_{\pi^+} - m_{e}) =139\,$MeV/$c^{2}$ can be produced in 
$\pi^{+}\rightarrow e^{+} N$ decays. 
The characteristics of HNL production in leptonic decays are detailed in~\cite{shrock1980new}. The decay rate depends on the HNL mass $m_{N}$ and the mixing parameter $|U_{e4}|^{2}$ as
\begin{equation}
\label{eq:branching}
 {\cal B}(\pi^+\to e^+N) = {\cal B}(\pi^+\to e^+\nu_e)\cdot\rho(m_N)\cdot|U_{e4}|^2,
\end{equation}
where ${\cal B}(\pi^+\to e^+\nu_e)$ is the measured branching fraction of the
SM leptonic decay and $\rho(m_N)$ is a kinematic factor:
\begin{equation}
\label{eq:rho}
\rho(m_N)= \frac{(x+y)-(x-y)^2}{x(1-x)^2}\cdot\lambda^{1/2}(1,x,y),
\end{equation}
with $x=( m_e/m_{\pi^+})^2$, $y=(m_N/m_{\pi^+})^2$ and 
$\lambda(a,b,c)=a^2+b^2+c^2-2(ab+bc+ac)$.
The product ${\cal B}(\pi^+\to e^+\nu_e)\cdot\rho(m_N)$ is of ${\cal O}(1)$
over most of the allowed $m_N$ range,  
drops to zero at the kinematic limit $m_N=(m_{\pi^+}-m_e)$ and reduces to 
${\cal B}(\pi^+\to e^+\nu_{e}) = 1.230(4)\times10^{-4}$~\cite{ParticleDataGroup:2024cfk}
for $m_N\to0$, as $\rho(0)=1$.

Searches for HNL production in $K^+\to e^+N$ and $K^+\to\mu^+N$
decays were reported by the NA62 experiment at CERN using the 2016--2018 dataset~\cite{NA62:2020mcv,NA62:2021bji}, which established 
90\,\% CL upper limits for $|U_{e4}|^2$ and $|U_{\mu4}|^2$ 
of ${\cal O}(10^{-9})$ and ${\cal O}(10^{-8})$, respectively, in a mass range of a few hundred MeV$/c^2$.
In a search for $\pi^+\to e^+N$ and $\pi^+\to \mu^+ N$ decays at rest, the PIENU experiment reported upper limits for $|U_{e4}|^2 < 10^{-7}$ and $|U_{\mu4}|^2 < 10^{-5}$ in the HNL mass ranges 60--125 MeV/c$^2$\cite{PIENU:2017wbj} and 16--33 MeV/c$^2$ \cite{pienuMu}, respectively. 

A search for $\pi^+\to e^+N$ decays in the HNL mass range 95--126
\,MeV$/c^2$ using the data collected by NA62 in 2017--2024 is reported here. The results,
assuming that the
HNL lifetime exceeds 50\,ns, are presented as upper limits for $|U_{e4}|^2$
at 90\,\% CL for a set of mass hypotheses. The NA62 experiment is not sensitive to $\pi^+\to \mu^+N$ in-flight decays.

\section{Beam, detector and data samples}
\label{sect:beam}
A description of the NA62 beamline and detector is given in~\cite{NA62:2017rwk}. An unseparated secondary beam of
$\pi^+$ (70\,\%), protons (23\,\%) and $K^+$ (6\,\%) is created by directing $400\,\mbox{GeV}/c$
protons extracted from the CERN SPS onto a beryllium target in
spills of 4.8\,s duration. The central beam momentum is
$75\,\mbox{GeV}/c$, with a momentum spread of 1\,\% (rms).
 
\begin{figure}
    \centering
    \includegraphics[width=1.0\textwidth]{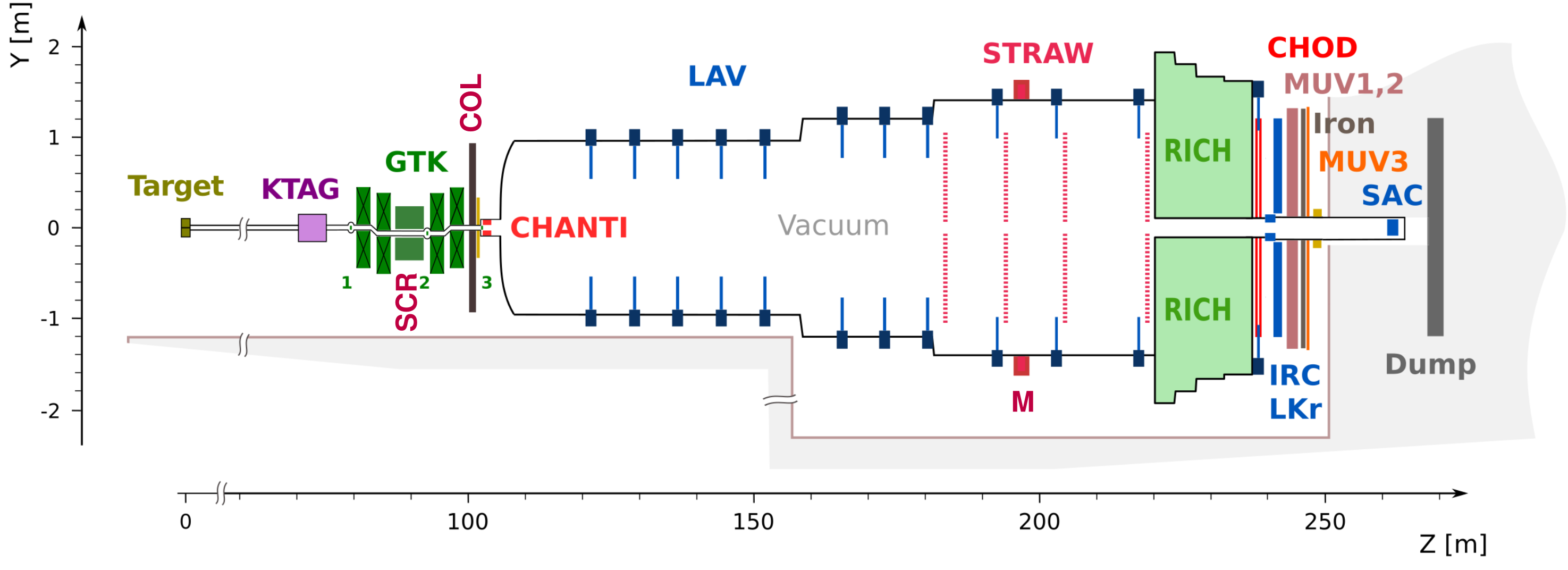}
    \caption{Schematic side view of the NA62 detector during the 2017--2018 data taking period.
    }
    \label{fig:det}
\end{figure}

 The detector layout is shown
schematically in Fig.~\ref{fig:det}. 
Beam kaons are tagged with 70\,ps time resolution by a differential Cherenkov counter (KTAG)~\cite{NA62:2023mud}.
A beam spectrometer (Gigatracker, GTK) provides measurement of momentum, direction, and time of beam particles with resolutions of 0.15\,GeV/$c$, 16\,$\mu$rad, and 100\,ps, respectively. The GTK is composed of silicon pixel matrices (with a pixel size of $300 \times 300$ 
$\mu$m$^2$) arranged in three stations in 2016--2018 and 
(after upgrades during the accelerator shutdown)
four stations in 2021--2024, along with two pairs of dipole magnets forming an achromat.
A toroidal muon scraper magnet (SCR)
is installed between GTK1 and GTK2. A 1.2\,m thick steel collimator (COL) with a central aperture of
$76\times40\,\mbox{mm}^2$ and outer
dimensions of $1.7\times1.8\,\mbox{m}^2$ is placed upstream of GTK3 to absorb
hadrons from upstream $K^+$ decays (a variable aperture collimator
of $0.15\times0.15\,\mbox{m}^2$ outer dimensions was used up to early 2018).
Inelastic interactions of beam particles in GTK3 are detected by an
array of scintillator hodoscopes (CHANTI) located immediately downstream of GTK3.
The beam is delivered into a vacuum tank evacuated to a pressure
of $10^{-6}\,\mbox{mbar}$, which contains a 75\,m long fiducial decay volume
(FV) starting 2.6\,m downstream of GTK3. 
The angular spread of the beam
at the FV entrance is 0.11\,mrad (rms) in both horizontal and vertical
planes. Downstream of the FV, undecayed beam particles continue
their path in vacuum.

Momenta of charged particles produced by $K^+$ and $\pi^+$ decays in the FV
are measured by a magnetic spectrometer (STRAW) located in the
vacuum tank downstream of the FV. The spectrometer consists of
four tracking chambers made of straw tubes and a dipole magnet
(M), located between the second and third chambers, which provides a
horizontal momentum kick of $270\,\mbox{MeV}/c$. The momentum
resolution achieved is $\sigma_p/p=(0.30\oplus0.005\cdot p)\,\%$, where the momentum $p$ is expressed in GeV/$c$.

A ring-imaging Cherenkov detector (RICH), consisting of a
17.5\,m long vessel filled with neon at atmospheric pressure (with
a Cherenkov threshold for muons of 9.5\,GeV/$c$), is used for the
identification of charged particles and time measurements with
70\,ps precision. Two scintillator hodoscopes (CHOD),
which include a matrix of tiles and two orthogonal planes of slabs,
arranged in four quadrants, provide trigger signals and time measurements with 200\,ps precision.

A $27\,X_0$ thick quasi-homogeneous liquid krypton (LKr) electromagnetic calorimeter is used for particle identification and photon detection. The calorimeter has an active volume of 
$7\,\mbox{m}^3$, is
segmented in the transverse plane into 13248~projective cells
of approximately
$2\times2\,\mbox{cm}^2$, and provides an energy resolution
$\sigma_E/ E = ( 4 .8/\sqrt{E} \oplus 11/E\oplus 0.9)\,\%$, where the energy $E$ is expressed in GeV.
To achieve hermetic acceptance for photons emitted in decays in
the FV at angles up to 50\,mrad to the beam axis, the LKr calorimeter is supplemented by annular lead glass detectors (LAV) installed
in 12~positions inside and downstream of the vacuum tank, and two
lead/scintillator sampling calorimeters (IRC, SAC) located close to
the beam axis. An iron/scintillator sampling hadronic calorimeter
formed of two modules (MUV1,2) and a muon detector (MUV3)
consisting of 148 scintillator tiles located behind an 80 cm thick
iron wall are used for particle identification.

The data samples used for the analysis  were
recorded in 2017--2018 and 2021--2024,
at a typical beam intensity of (2--4)$\times10^{12}$ protons per spill corresponding to (330--580)\,$\times10^6$ beam particles per second at the FV entrance,
and a mean $K^+$ and $\pi^+$ decay rate in the FV of few $10^6/$s.
The trigger system  consists of a hardware low level (L0) and a software high level (L1)~\cite{cortina2023performance}. The main trigger line dedicated to the $K^+\to\pi^+\nu\bar\nu$ measurement~\cite{NA62:2024pjp}
is used for this analysis. 
The L0 trigger requires the following conditions: at least two signals in the RICH; at least one signal in any quadrant, no signals in diagonally-opposite quadrants and fewer than five signals in the CHOD; no signals in MUV3; between 5\,GeV and 30\,GeV (40\,GeV) energy deposited in the LKr for 2017--2018 (2021--2024) data. 
The L1 trigger requires the following conditions: a kaon signal in the KTAG  within 5~ns of the L0 trigger time;
a positively charged STRAW track with momentum below 50 GeV/$c$ 
compatible with originating from a beam particle decay in the FV;
fewer than two signals in LAV2--11.
About 1/3 of the beam pions are accepted by the L1-KTAG condition because of an accidental time-coincidence with a beam kaon.

Monte Carlo (MC) simulations of particle interactions with the detector and its response are performed using a software package based on the Geant4 toolkit~\cite{GEANT4:2002zbu}.
In addition, accidental activity is simulated and the response of the L0 trigger conditions is emulated.

\section{Event selection, normalization and background}
\label{sec:selection}

The $\pi^+\to e^+N$ decay is characterized by a single positron in the final
state, as in the SM $\pi^+\to e^+\nu_e$ decay. 
Assuming a HNL lifetime greater than 50\,ns, and given 
that the HNLs produced in $\pi^{+}\rightarrow e^{+}N$ decays would be boosted 
by a Lorentz factor of ${\cal O}(500)$ for the considered mass range, the decays of HNLs into SM particles 
in the detector
can be neglected.
The principal selection criteria follow:

\begin{itemize}
\item A positively charged track reconstructed in the STRAW spectrometer
with momentum in the range 5--30\,$\mbox{GeV}/c$ is required, consistent with the L0 LKr energy trigger condition. The different L0 LKr trigger conditions in 2017--2018 and 2021--2024 only affect the selection acceptance for HNL masses below 95\,MeV$/c^2$. The particle  trajectory through the STRAW chambers
and its extrapolations to the LKr calorimeter and RICH front planes
should be within the geometrical acceptance of these
detectors. The track time is evaluated as the mean time of
the RICH signals spatially associated with the track.
\item The following two particle identification criteria are applied to select a positron: the ratio of energy $E$, associated with the extrapolated track in the LKr calorimeter, to momentum $p$, measured by
the STRAW spectrometer, is required to be $0.9 < E / p < 1.1$;
a particle identification algorithm~\cite{Muller:1993ig}  based on the RICH signal
pattern within 3\,ns of the track time is used.
\item
A matching beam track in the GTK is found considering the time difference, $\Delta t$, and spatial compatibility quantified by the closest distance of approach, CDA, between a GTK track and the STRAW track identified as a positron. 
A discriminant
${\cal D}(\Delta t,$ CDA$)$ is defined using the $\Delta t$ and CDA distributions obtained from reconstructed 
$K^+\to\pi^+\pi^+\pi^-$ decays in data. 
Among GTK tracks with
$|\Delta t| < 0.5$\,ns, the track with the largest
discriminant value is considered as the parent beam particle. 
Additionally, CDA $< 3\,\mbox{mm}$ is required to reduce background from
$\pi^+\to\mu^+\nu_\mu$ and $K^+\to\mu^+\nu_\mu$ decays in the FV followed by muon
decay $\mu^+\to e^+\nu_e\bar\nu_\mu$. The decay vertex is defined as the midpoint of the segment 
of closest approach of the GTK and the identified positron tracks, taking into
account the residual magnetic field in the vacuum tank.
The reconstructed decay vertex of the $\pi^+$ ($K^+$)  must be located within the FV and at least 5 meters downstream from its origin.
This condition suppresses
background from $\pi^+\to\mu^+\nu_\mu$ and $K^+\to\mu^+\nu_\mu$ decays upstream of GTK3 followed by muon decays in the FV.
\item
 The positron track must not form a vertex
with any other STRAW track. 
No LKr energy deposit is allowed within 4\,ns of the positron track time and not spatially associated with it.
No activity in the large-angle
(LAV) and small-angle (SAC, IRC) photon veto detectors within
5\,ns of the track time is allowed. These conditions suppress backgrounds from multi-body $K^+$ decays.
\end{itemize}

\begin{figure}
    \includegraphics[width=0.49\textwidth]{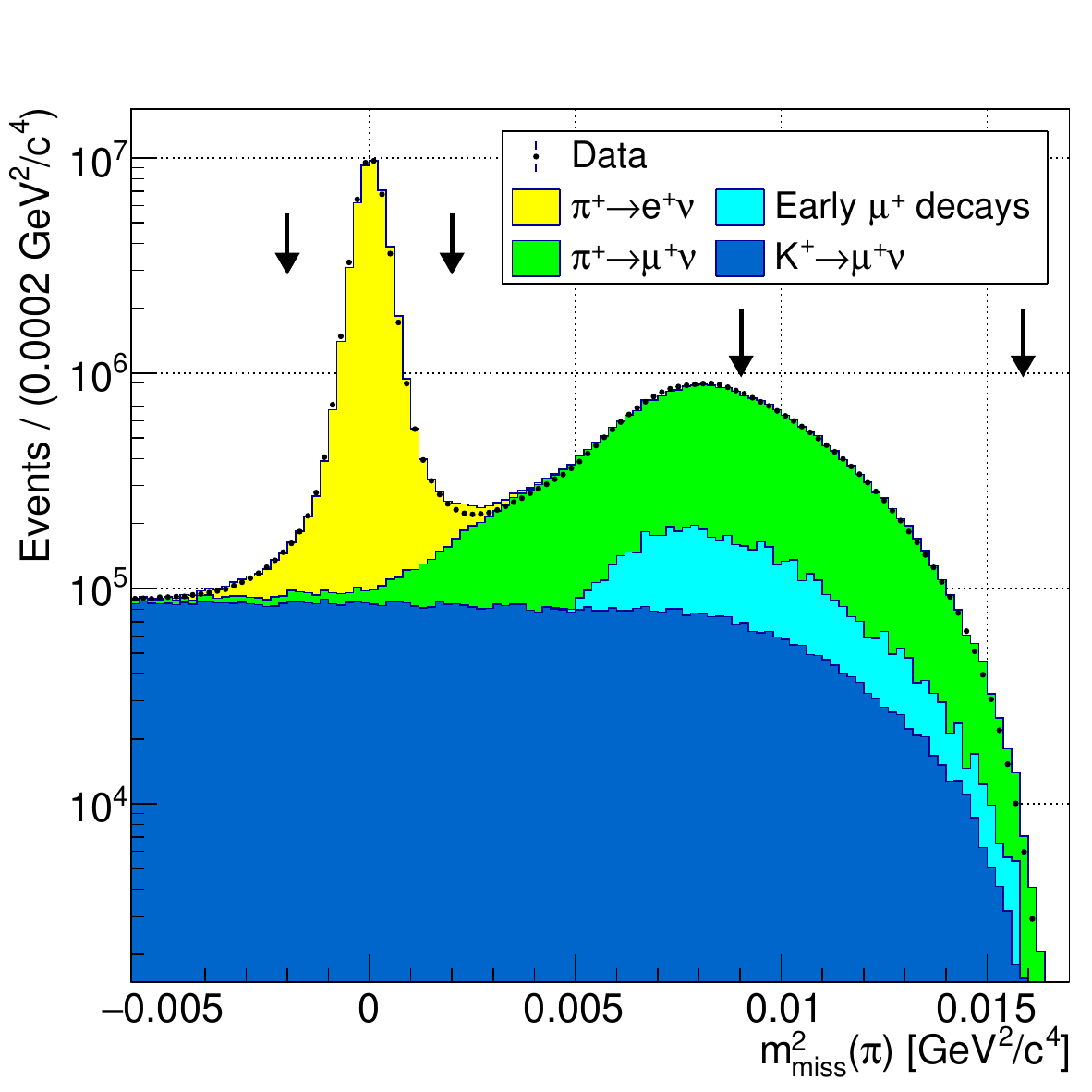}
    \hfill
    \includegraphics[width=0.49\textwidth]{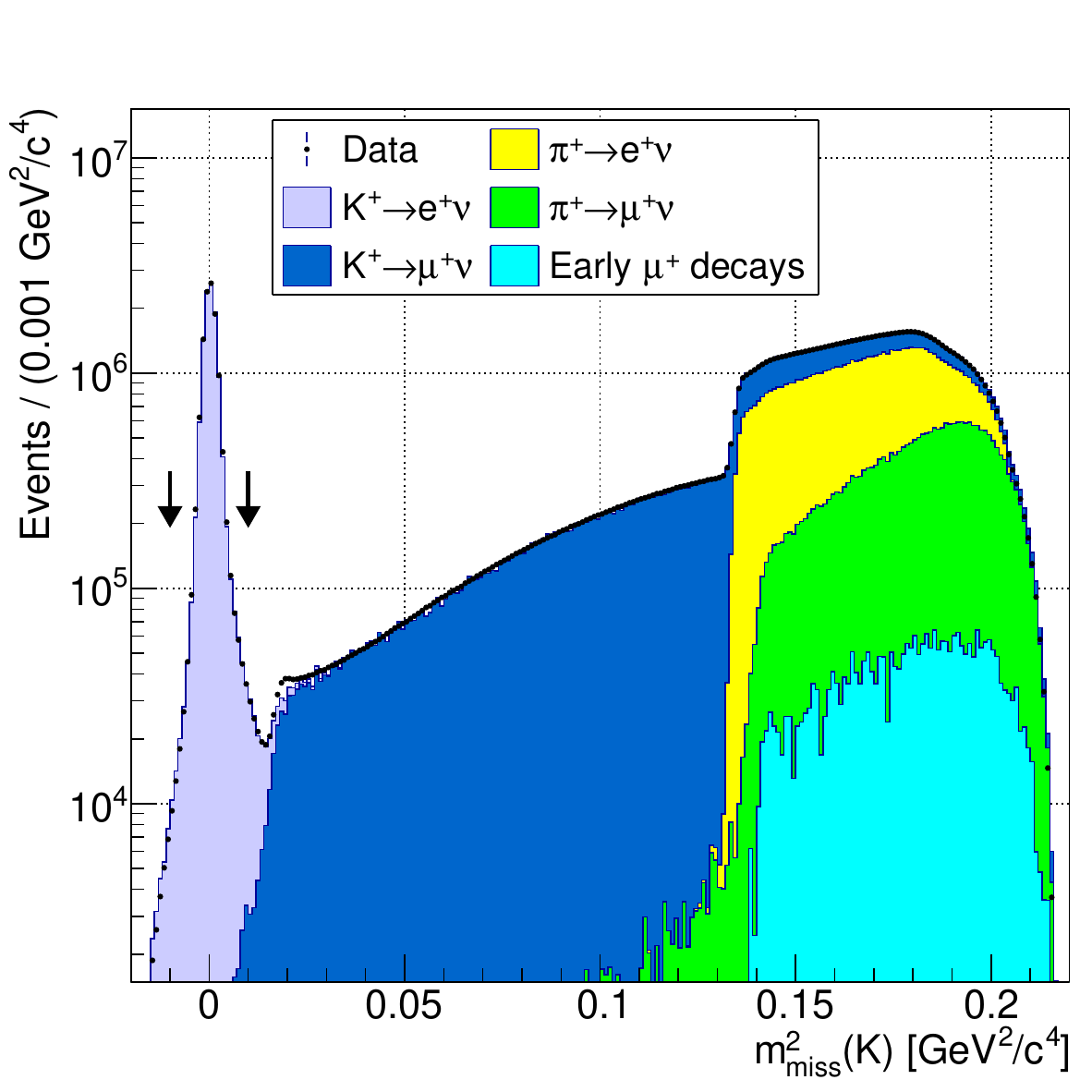}
\caption{Squared missing mass distributions assuming the beam particle is a $\pi^+$ (left) and a $K^+$ (right). 
Pairs of vertical arrows indicate the squared missing mass ranges used for the selection of $\pi^+\to e^+\nu_e$ decays and HNL search region (left), and for the selection of $K^+\to e^+\nu_e$ decays (right). Kaon and pion decays to muons contribute via muon decays in flight.}
\label{fig:mmiss}
\end{figure}

The squared missing mass is computed assuming that the beam particle is a pion or a kaon, respectively, 
as $m_{\rm miss}^2(\pi) = (P_\pi - P_e)^2$
and $m_{\rm miss}^2(K) = (P_K - P_e)^2$,
where $P_\pi$, $P_K$ and $P_e$ are the pion, kaon, and positron 4-momenta, obtained
from the 3-momenta measured by the GTK and STRAW detectors
and using the $\pi^+$, $K^+$ and $e^+$ mass values \cite{ParticleDataGroup:2024cfk}.
The $m_{\rm miss}^2$ spectra of the events
selected from data and simulated samples are displayed in Fig.~\ref{fig:mmiss}.
Event candidates from the SM leptonic decay
$\pi^+\to e^+\nu_e$ ($K^+\to e^+\nu_e$)
are observed as peaks at $m_{\rm miss}^2 = 0$ with resolutions of $3.3\times10^{-4} \,(1.5\times10^{-3})\,\mbox{GeV}^2/c^4$.
The SM normalization regions are defined as $|m_{\rm miss}^2(\pi)|<0.002\,\mbox{GeV}^2/c^4$ for the pion case and $|m_{\rm miss}^2(K)|<0.01\,\mbox{GeV}^2/c^4$ for the kaon case.
Backgrounds from $\pi^{+}\rightarrow\mu^{+}\nu_{\mu}$ and $K^{+}\rightarrow\mu^{+}\nu_{\mu}$ decays contribute mainly via
muon decays in flight.
Only a small fraction of the background from $K^{+}\rightarrow\mu^{+}\nu_{\mu}$ decays followed by muon decay contributes with positive 
values of $m_{\rm miss}^2(\pi)$, while $K^{+}\rightarrow e^+\nu_e$ decays contribute with values smaller than $-0.05\,\mbox{GeV}^2/c^4$.

The $K^+$ decay component in the $m_{\rm miss}^2(\pi)$ spectrum is obtained from simulations and scaled to the number $N_K$ of $K^+$ decays observed in the FV.  This number is evaluated using the number of $K^+\to e^+\nu_e$ candidates reconstructed in data within the SM normalization region (Fig.~\ref{fig:mmiss}-right). 
The dominant background, $K^+\to\mu^+\nu_\mu$
decay followed by muon decay, 
contributes at the negligible level of 
0.1\,\%.
For each of the two data-taking periods,  
the number of $K^+$ decays is computed as 
\begin{equation}
N_K = \frac{N_{\rm SM}^{K}}{A_e^{K}\cdot{\cal B}(K^+\to e^+\nu_e)},   
\end{equation}
where $N_{\rm SM}^{K}$ 
is the number of selected data events in the SM normalization region; $A_e^{K}$ is the acceptance 
of the selection for
the $K^+\to e^+\nu_e$ decay 
evaluated with simulations; 
and ${\cal B}(K^+\to e^+\nu_e)$ 
is the branching fraction of this decay~\cite{ParticleDataGroup:2024cfk}. 
Data losses due to L0 trigger inefficiencies and accidental vetoes are included in the $N_K$  definition, which makes the value of $N_K$ specific to this analysis.

The number $N_\pi$ of $\pi^+$ decays in the FV is evaluated 
from the number of $\pi^+\to e^+\nu_e$ data candidates  reconstructed in the SM normalization region (Fig.~\ref{fig:mmiss}-left), subtracting the number of background events, $N_{\rm K\mu2}^{\pi}$, originating 
from $K^+\to\mu^+\nu_\mu$ decays: 
\begin{equation}
N_\pi = \frac{N_{\rm SM}^{\pi} - N_{\rm K\mu2}^{\pi}}
{A_e^{\pi}\cdot{\cal B}(\pi^+\to e^+\nu_e) +  
A_\mu^{\pi}\cdot{\cal B}(\pi^+\to\mu^+ \nu_\mu)},
\end{equation}
where $N_{\rm SM}^{\pi}$  and $N_{\rm K\mu2}^{\pi}$ are the numbers of selected data and background events in the SM normalization region; 
$A_e^{\pi}$ and $A_\mu^{\pi}$ are the acceptances of the selection for
the $\pi^+\to e^+\nu_e$ decay and the $\pi^+\to\mu^+\nu_\mu$ decay followed by muon decay, respectively, 
evaluated with simulations;  
and ${\cal B}(\pi^+\to e^+\nu_e)$  and ${\cal B}(\pi^+\to\mu^+\nu_\mu)$ are the branching fractions of these decays~\cite{ParticleDataGroup:2024cfk}.
The number of background events in the SM normalization region is 
\begin{equation}
N_{\rm K\mu2}^{\pi} = N_{K}\cdot A_{K\mu2}^{\pi} \cdot {\cal B}(K^+\to\mu^+\nu_\mu) ,
\end{equation}
 where  $A_{K\mu2}^{\pi}$ is the acceptance of the $\pi^+\to e^+\nu_e$ selection for the $K^+\to\mu^+\nu_\mu$ decay followed by
muon decay, evaluated with simulations,  and ${\cal B}(K^+\to\mu^+\nu_\mu)$ is the corresponding branching fraction.

  The numerical values of quantities involved in the evaluations above and their uncertainties are listed in Table~\ref{tab:numbers}.
The obtained $N_{K}/N_{\pi}$ ratio is consistent with the beam composition taking into account the Lorentz factors, $K^+$ and $\pi^+$ lifetimes and L1-KTAG efficiency for beam pions.
The simulated normalization and background sources describe the data within a few percent over the whole missing mass range.
The uncertainties in $N_K$ and $N_\pi$
include contributions from the statistical accuracy of the simulation and the uncertainty of the external branching fractions. The systematic uncertainties are estimated by varying the width of the SM normalization regions.

\begin{table}
\centering
\caption{Values used to obtain the numbers of kaon and pions decays in the FV.}
\label{tab:numbers}
\vspace{10pt} 
\begin{tabular}{llcc}  
\multicolumn{2}{l}{Period} & 2017--2018 & 2021--2024 \\
\hline
\hline
$N_{\rm SM}^{K}$ & $[10^6]$              & $3.65 $  & $7.66 $ \\
$A_e^{K}$ &  $[10^{-2}]$                      & \phantom{9}$3.962 (6) $ & \phantom{19}$3.667 (6) $  \\
$N_K$ & $[10^{12}]$                            & $5.82(3)$   &  $13.19(7)$\\
\hline
Total $N_K$ & $[10^{12}]$          & \multicolumn{2}{c}{$19.02(8) $}\\
\hline 
\hline
$N_{\rm SM}^{\pi}$  &$[10^{7}]$ & $1.35$   & $3.35$ \\
$A_e^{\pi}$ & $[10^{-2}]$     &  $6.070(8)$ & $5.394(7)$  \\
$A_\mu^{\pi}$  &$[10^{-8}]$   & $6.89(14)$  & $7.90(15)$ \\
$A_{K\mu2}^{\pi}$& $[10^{-7}]$ & $1.408(4)$          &  $1.410(5)$\\
$N_{\rm K\mu2}^{\pi}$ & $[10^{5}]$   & $5.21(4)$ &  $11.8 (1)$ \\
$N_\pi$  &  $[10^{12}]$                 & $1.72(1)$ & $4.81(3)$\\
\hline
Total $N_\pi$ &$[10^{12}]$  & \multicolumn{2}{c}{$6.54(3)$} \\
\hline
\hline
  \end{tabular}
  \end{table}

\section{Search procedure}
\label{sec:search}
A peak-search procedure measures the $\pi^+\to e^+N$ decay rate
with respect to the $\pi^+\to e^+\nu_e$ decay rate for a set of HNL mass hypotheses $m_N$.  
This approach benefits from first-order cancellations of residual detector inefficiencies
common to signal and normalization modes and not fully accounted for in simulations.
The expected number of $\pi^+\to e^+N$
signal events, $N_S$, can be written as
\begin{equation}
\label{eq:signal}
N_S = {\cal B}(\pi^+\to e^+N)/{\cal B}_{\rm SES}(\pi^+\to e^+N)=|U_{e4}|^2/|U_{e4}|_{\rm SES}^2,
\end{equation}
where the single event sensitivity ${\cal B}_{\rm SES}(\pi^+\to e^+N)$ and the
mixing parameter $|U_{e4}|_{\rm SES}^2$ corresponding to the expectation of one signal
event are defined as
\begin{equation}
\label{eq:ses}
{\cal B}_{\rm SES}(\pi^+\to e^+N) = \frac{1}{N_\pi \cdot A_N} {\rm ~~~~~and~~~~~} 
|U_{e4}|_{\rm SES}^2 = \frac{{\cal B}_{\rm SES}(\pi^+\to e^+N)}{{\cal B}(\pi^+\to e^+\nu_e)\cdot\rho(m_N)},
\end{equation}
where
$A_N$ is the signal
selection acceptance for decays in the FV
and $\rho(m_N)$ is the kinematic factor defined in Eq.~\ref{eq:rho}.

\begin{figure}
    \includegraphics[width=0.49\textwidth]{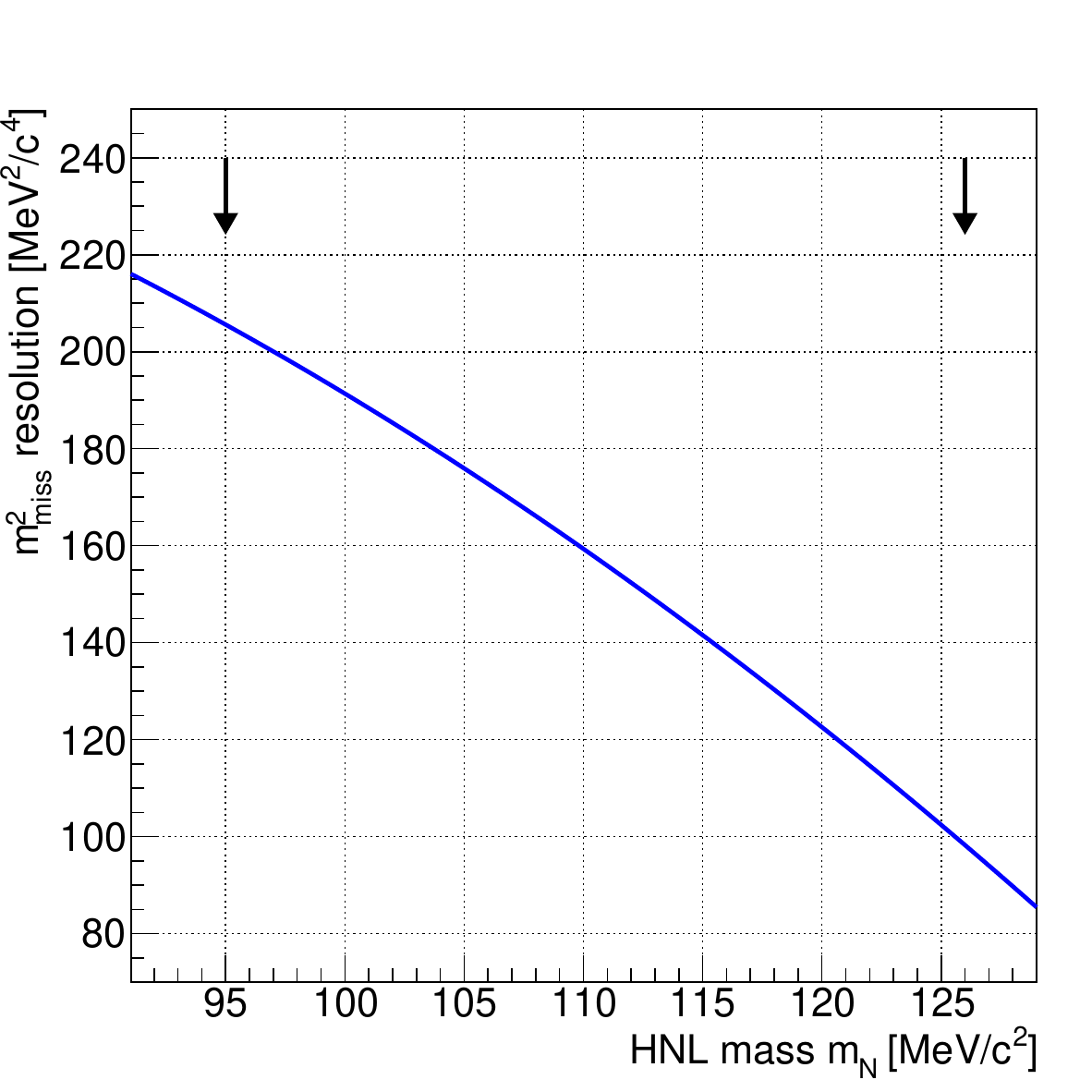}
    \hfill
    \includegraphics[width=0.49\textwidth]{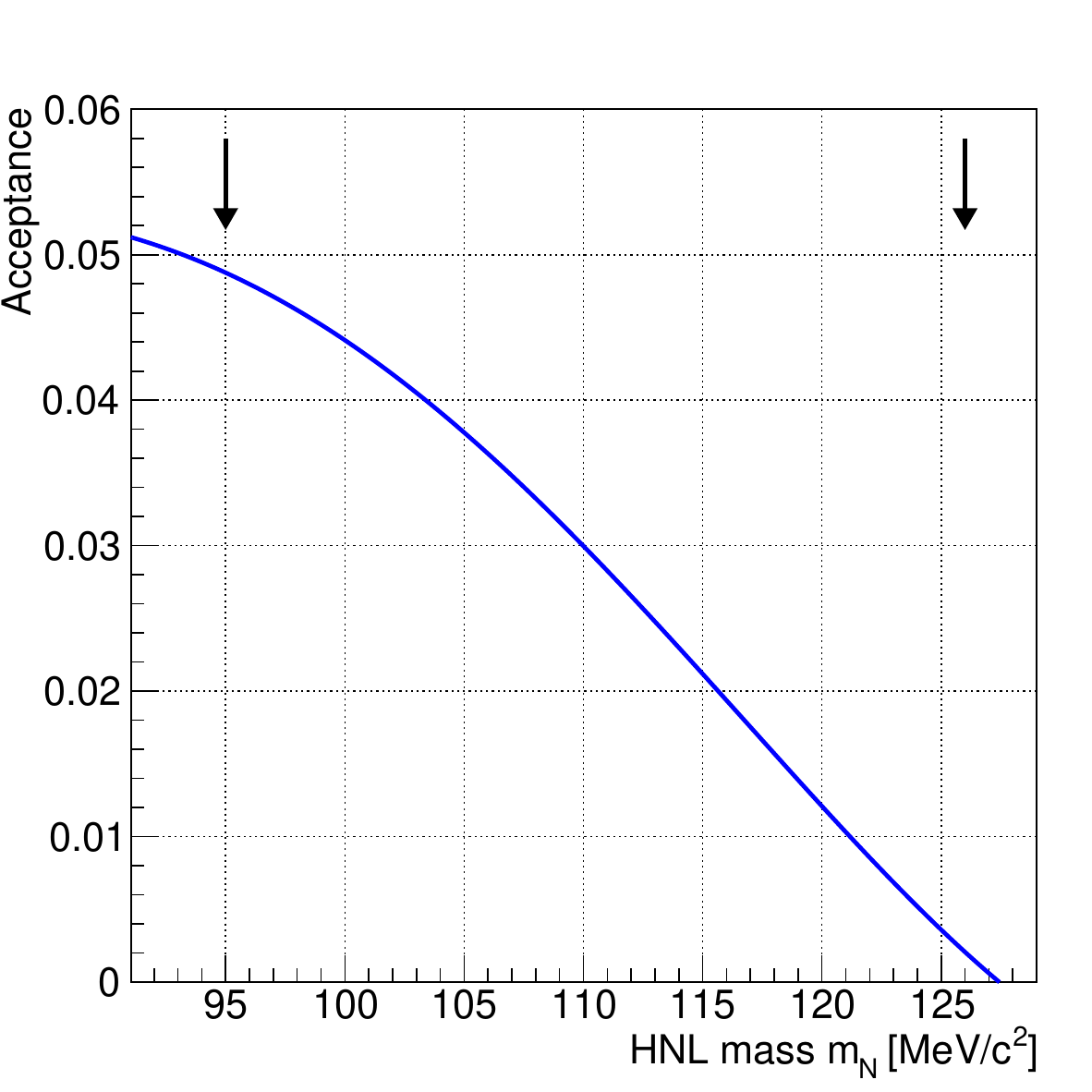}
\caption{Squared missing mass resolution $\sigma_{m^2}$ (left) and acceptance $A_{N}$ (right) evaluated from simulations as functions of HNL mass. The boundaries of the HNL search region are indicated by the vertical arrows.}
\label{fig:resolution}
\end{figure}

The search uses a data-driven estimation of
the background to $\pi^+\to e^+N$ decays, which is valid in the absence of peaking signal-like background structures in the squared missing mass spectrum. 
The $\pi^+\to e^+N$ process is investigated in 63~mass hypotheses,
$m_N$, within the HNL search region 95--126\,$\mbox{MeV}/c^2$.
The distance between adjacent mass hypotheses is equal to 3/4 of the local
mass resolution $\sigma_{m^2}$ shown in Fig.~\ref{fig:resolution}-left.
The number of observed events, $N_{\rm obs}$, is counted in the interval 
$|m_{\rm miss}^2(\pi) - m_N^2| < 1.5\,\sigma_{m^2}$
for each mass hypothesis.
Sidebands are defined as $1.5\,\sigma_{m^2} < | m_{\rm miss}^2(\pi) - m_N^{2} | < 9\,\sigma_{m^2}$,
requiring missing mass values in
the range 92.2--127.6\,MeV/$c^2$ to take into account the
search region boundaries. The number of expected background
events, $N_{\rm exp}$, within each signal window is evaluated using a third-order polynomial fit to the sidebands of the $m^2_{\rm miss}(\pi)$ spectrum.
For mass values below 95 MeV/c$^{2}$ the number of expected background events cannot be satisfactorily determined with the described method.
The uncertainty, $\delta N_{\rm exp}$,
in the number of expected background events includes statistical
and systematic components. The former comes from the statistical errors in the fit parameters, and the latter is evaluated as the
difference between $N_{\rm exp}$ values obtained from fits using third and fourth 
order polynomials, and the 
differences in $N_{\rm exp}$ from sidebands defined as 
$8.25\,\sigma_{m^2}$ and $9.75\,\sigma_{m^2}$ with respect to the original definition.  
The dominant contribution to $\delta N_{\rm exp}$ is statistical, except near the boundaries of the HNL search region where
the systematic uncertainty is comparable.

\begin{figure}
\includegraphics[width=0.49\textwidth]{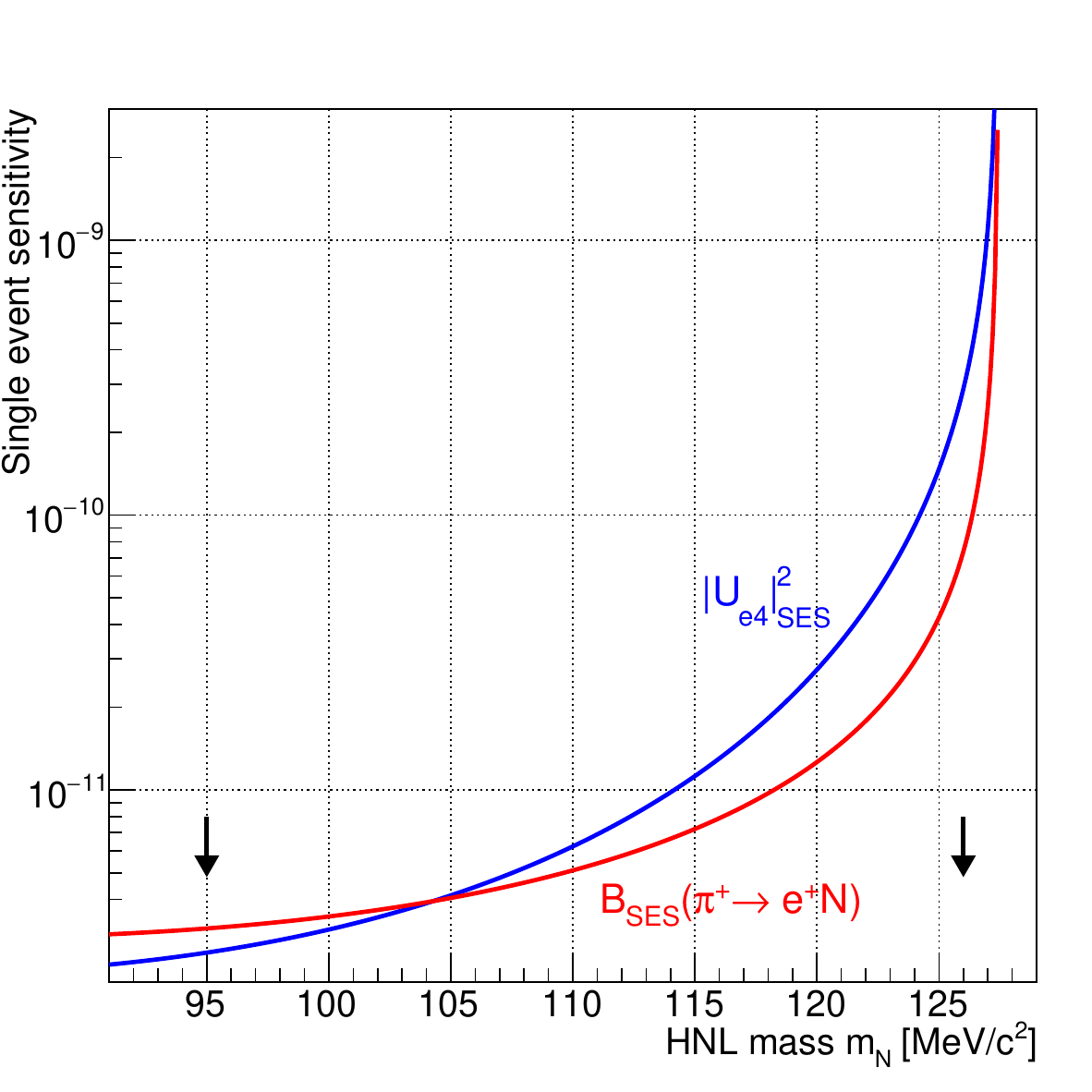}
\hfill
\includegraphics[width=0.49\textwidth]{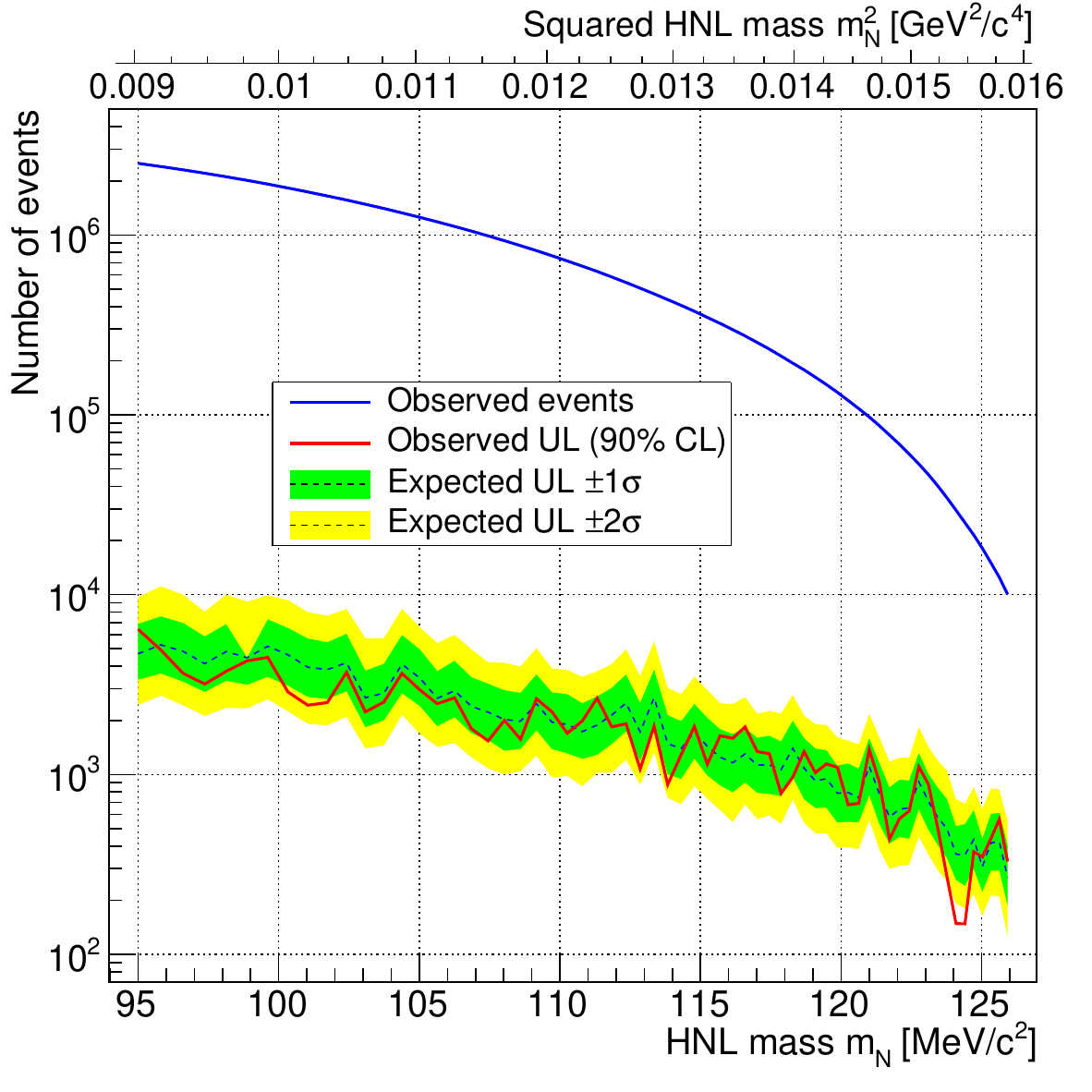}
\caption{Left: single event sensitivities ${\cal B}_{\rm SES}(\pi^+\to e^+N)$ and $|U_{e4}|_{\rm SES}^2$ as functions of the HNL mass, evaluated according to Eq.~\ref{eq:ses}. Right: number of observed events $N_{\rm obs}$, observed and expected upper limits at 90\,\% CL for the number of $\pi^{+}\rightarrow e^{+} N$ events, and the expected $\pm$1$\sigma$ and $\pm$2$\sigma$ bands in the background-only hypothesis for each HNL mass value considered.}
\label{fig:ses}
\end{figure}
\begin{figure}
\includegraphics[width=0.49\textwidth]{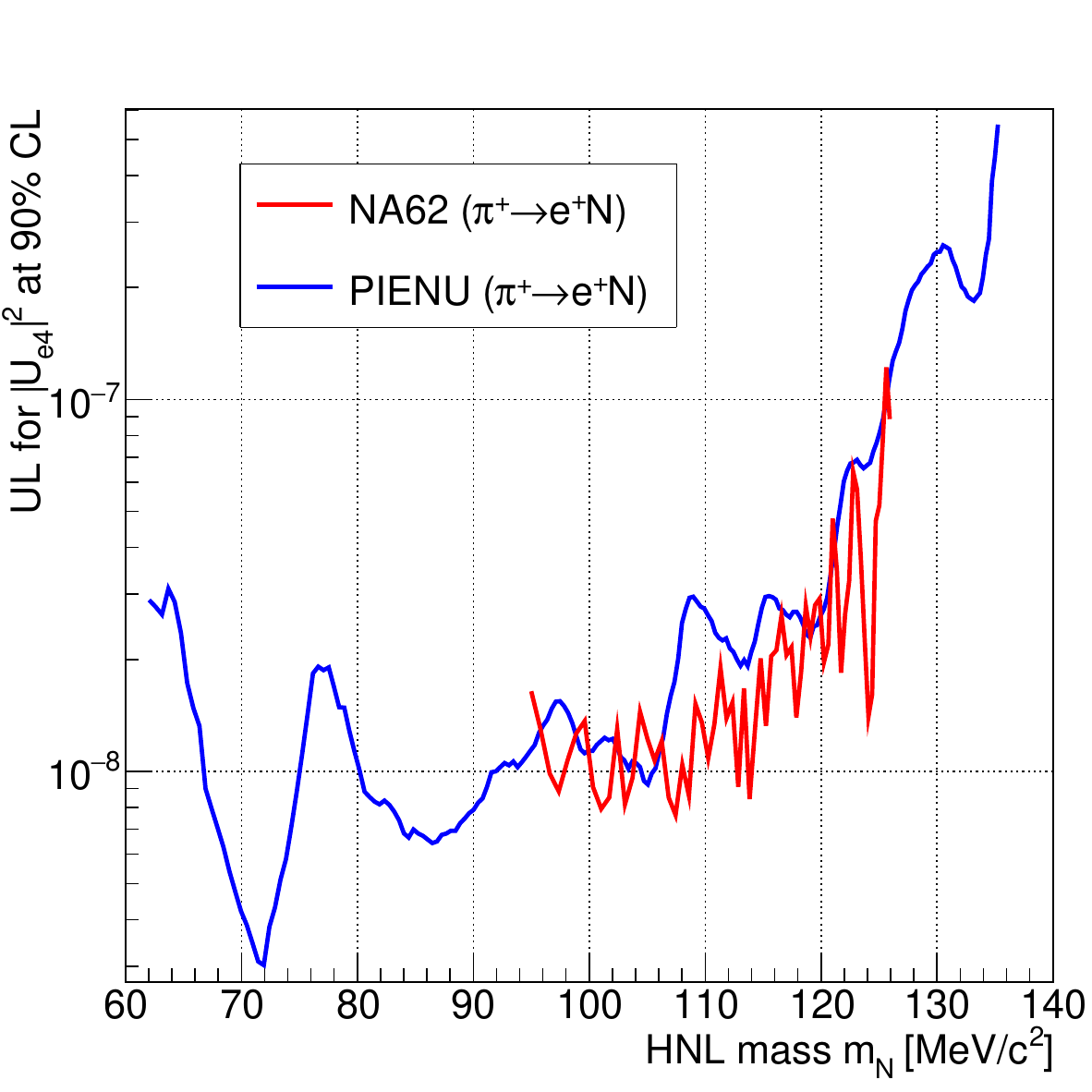}
\hfill
\includegraphics[width=0.49\textwidth]{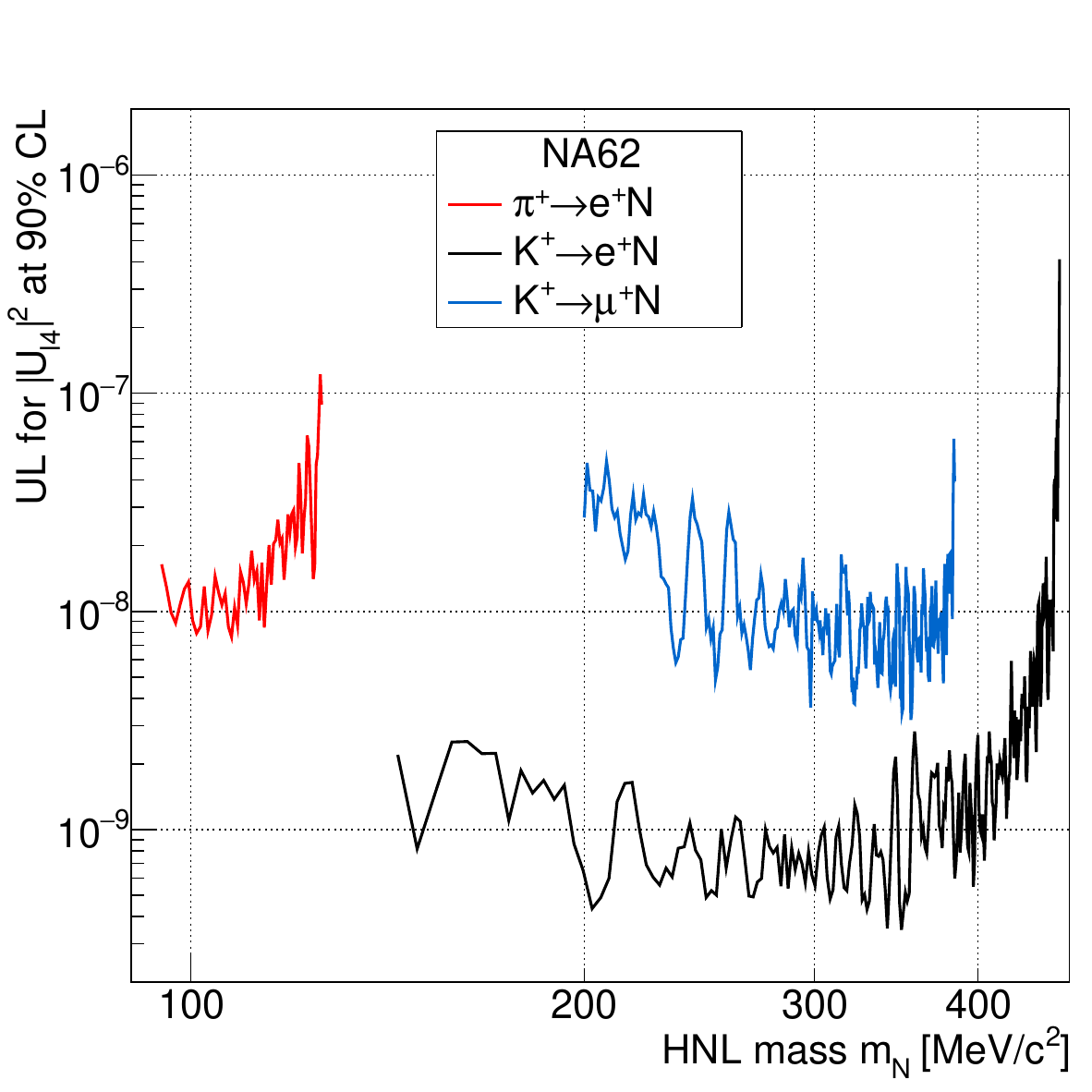}
\caption{Left: upper limits obtained by NA62 at 90\,\% CL for $|U_{e4}|^2$ and upper limits established by the 
PIENU experiment in a wider mass range \cite{PIENU:2017wbj}. 
Right: summary of upper limits at 90\,\% CL for $|U_{e4}|^{2}$ and $|U_{\mu4}|^{2}$ obtained from HNL production searches at NA62 in kaon (2016--2018 data)~\cite{NA62:2020mcv,NA62:2021bji} and pion (2016--2024 data) decays.} 
\label{fig:result}
\end{figure}

\section{Results}
\label{sec:results}
Upper limits at 90\,\% CL for the mixing parameter $|U_{e4}|^2$ are obtained according to 
Eqs.~\ref{eq:signal} and~\ref{eq:ses},
using the signal selection acceptance, $A_N(m_N)$,
obtained from simulations (Fig.~\ref{fig:resolution}-right), to compute the single event sensitivities
${\cal B}_{\rm SES}(\pi^+\to e^+N)$ and $|U_{e4}|_{\rm SES}^2$
(Fig.~\ref{fig:ses}-left). The quantities $N_{\rm obs}$, $N_{\rm exp}$, and $\delta N_{\rm exp}$ are used to evaluate the
upper limit at 90\,\% CL for the number of $\pi^+\to e^+N$ decays, $N_S$, for each HNL mass hypothesis using the CL$_S$ method~\cite{Read:2002hq}.
The values of $N_{\rm obs}$, the upper limits obtained for $N_S$, and the expected $\pm1\,\sigma$
and $\pm2\,\sigma$ bands of variation of $N_S$ in the background-only hypothesis are shown in Fig.~\ref{fig:ses}-right.
 
The upper limits obtained for $|U_{e4}|^2$ are displayed in
Fig.~\ref{fig:result}-left, together with those published by the PIENU experiment~\cite{PIENU:2017wbj}.
These results are  
similar in the range considered, 
with different experimental techniques employed,
namely decay in flight and stopped pions. Fig.~\ref{fig:result}-right shows the NA62 results obtained in searches for HNL production in $K^+ \to e^+ N$ \cite{NA62:2020mcv}, $K^+ \to \mu^+ N$ \cite{NA62:2021bji} and $\pi^+ \to e^+ N$ decays.

\section{Summary}
\label{summary}
A search for HNL production in $\pi^+\to e^+N$ decays has been
performed using the data collected by the NA62 experiment
at CERN in 2017--2024. Upper limits for the mixing parameter
$|U_{e4}|^2$ have been established at the
$10^{-8}$ level over the HNL mass range 95--126\,$\mbox{MeV}/c^2$ assuming the mean lifetime exceeds 50\,ns.
These limits are comparable to, or more stringent than, those obtained in \cite{PIENU:2017wbj} for the mass region covered in this study.
An improved  sensitivity of this analysis in terms of $|U_{e4}|^2$
is expected including future NA62 data.

\section*{Acknowledgments}

It is a pleasure to express our appreciation to the staff of the CERN laboratory and the technical
staff of the participating laboratories and universities for their efforts in the operation of the
experiment and data processing.

The cost of the experiment and its auxiliary systems was supported by the funding agencies of 
the Collaboration Institutes. We are particularly indebted to: 
F.R.S.-FNRS (Fonds de la Recherche Scientifique - FNRS), under Grants No. 4.4512.10, 1.B.258.20, Belgium;
CECI (Consortium des Equipements de Calcul Intensif), funded by the Fonds de la Recherche Scientifique de Belgique (F.R.S.-FNRS) under Grant No. 2.5020.11 and by the Walloon Region, Belgium;
NSERC (Natural Sciences and Engineering Research Council), funding SAPPJ-2018-0017,  Canada;
MEYS (Ministry of Education, Youth and Sports) funding LM 2018104, Czech Republic;
BMBF (Bundesministerium f\"{u}r Bildung und Forschung), Germany;
INFN  (Istituto Nazionale di Fisica Nucleare),  Italy;
MIUR (Ministero dell'Istruzione, dell'Universit\`a e della Ricerca),  Italy;
CONACyT  (Consejo Nacional de Ciencia y Tecnolog\'{i}a),  Mexico;
IFA (Institute of Atomic Physics) Romanian 
CERN-RO Nr. 06/03.01.2022
and Nucleus Programme PN 19 06 01 04,  Romania;
MESRS  (Ministry of Education, Science, Research and Sport), Slovakia; 
CERN (European Organization for Nuclear Research), Switzerland; 
STFC (Science and Technology Facilities Council), United Kingdom;
NSF (National Science Foundation) Award Numbers 1506088 and 1806430,  U.S.A.;
ERC (European Research Council)  ``UniversaLepto'' advanced grant 268062, ``KaonLepton'' starting grant 336581, Europe.

Individuals have received support from:
Charles University (grants UNCE 24/SCI/016, PRIMUS 23/SCI/025), 
Ministry of Education, Youth and Sports (project FORTE CZ.02.01.01/ \\
00/22-008/0004632), Czech Republic;
Czech Science Foundation (grant 23-06770S);  \\ 
Agence Nationale de la Recherche (grant ANR-19-CE31-0009), France;
Ministero dell'Istruzione, 
dell'Universit\`a e della Ricerca (MIUR  ``Futuro in ricerca 2012''  grant RBFR12JF2Z, Project GAP), Italy;
Nuclemedica Soluciones, San Luis Potos\'{i}, Mexico;
the Royal Society  (grants UF100308, UF0758946), United Kingdom;
STFC (Rutherford fellowships ST/J00412X/1, \\
ST/M005798/1), United Kingdom;
ERC (grants 268062,  336581 and  starting grant 802836 ``AxScale'');
EU Horizon 2020 (Marie Sk\l{}odowska-Curie grants 701386, 754496, 842407, 893101, 101023808).

 \bibliographystyle{elsarticle-num} 
 \bibliography{main}

@article{NA62:2020mcv,
    author = "Cortina Gil, Eduardo and others",
    collaboration = "NA62",
    title = "{Search for heavy neutral lepton production in $K^+$ decays to positrons}",
    eprint = "2005.09575",
    archivePrefix = "arXiv",
    primaryClass = "hep-ex",
    reportNumber = "CERN-EP-2020-089",
    doi = "10.1016/j.physletb.2020.135599",
    journal = "Phys. Lett. B",
    volume = "807",
    pages = "135599",
    year = "2020"
}

@article{PIENU:2017wbj,
    author = "Aguilar-Arevalo, A. and others",
    collaboration = "PIENU",
    title = "{Improved search for heavy neutrinos in the decay $\pi\rightarrow e\nu$}",
    eprint = "1712.03275",
    archivePrefix = "arXiv",
    primaryClass = "hep-ex",
    doi = "10.1103/PhysRevD.97.072012",
    journal = "Phys. Rev. D",
    volume = "97",
    pages = "072012",
    year = "2018"
}

@article{pienuMu,
title = {Search for heavy neutrinos in $\pi\rightarrow\mu\nu$ decay},
journal = {Phys. Lett. B},
volume = {798},
pages = {134980},
year = {2019},
issn = {0370-2693},
doi = {https://doi.org/10.1016/j.physletb.2019.134980},
author = {A. Aguilar-Arevalo and others}
}

@article{ParticleDataGroup:2024cfk,
    author = "Navas, S. and others",
    collaboration = "Particle Data Group",
    title = "{Review of particle physics}",
    doi = "10.1103/PhysRevD.110.030001",
    journal = "Phys. Rev. D",
    volume = "110",
    pages = "030001",
    year = "2024"
}

@article{NA62:2017rwk,
    author = "Cortina Gil, Eduardo and others",
    collaboration = "NA62",
    title = "{The Beam and detector of the NA62 experiment at CERN}",
    eprint = "1703.08501",
    archivePrefix = "arXiv",
    primaryClass = "physics.ins-det",
    doi = "10.1088/1748-0221/12/05/P05025",
    journal = "JINST",
    volume = "12",
    pages = "P05025",
    year = "2017"
}

@article{NA62:2021bji,
    author = "Cortina Gil, Eduardo and others",
    collaboration = "NA62",
    title = "{Search for $K^+$ decays to a muon and invisible particles}",
    eprint = "2101.12304",
    archivePrefix = "arXiv",
    primaryClass = "hep-ex",
    reportNumber = "CERN-EP-2021-018",
    doi = "10.1016/j.physletb.2021.136259",
    journal = "Phys. Lett. B",
    volume = "816",
    pages = "136259",
    year = "2021"
}

@article{NA62:2024pjp,
    author = "Cortina Gil, Eduardo and others",
    collaboration = "NA62",
    title = "{Observation of the K$^{+}$\textrightarrow{}$ {\pi}^{+}\nu \overline{\nu} $ decay and measurement of its branching ratio}",
    eprint = "2412.12015",
    archivePrefix = "arXiv",
    primaryClass = "hep-ex",
    doi = "10.1007/JHEP02(2025)191",
    journal = "JHEP",
    volume = "02",
    pages = "191",
    year = "2025"
}

@article{GEANT4:2002zbu,
    author = "J. Allison and others",
    collaboration = "GEANT4",
    title = "{Recent developments in Geant4}",
    doi = "10.1016/j.nima.2016.06.125",
    journal = "Nucl. Instrum. Meth. A",
    volume = "835",
    pages = "186",
    year = "2016"
}

@article{Read:2002hq,
    author = "Read, Alexander L.",
    editor = "Whalley, M. R. and Lyons, L.",
    title = "{Presentation of search results: The $CL_s$ technique}",
    doi = "10.1088/0954-3899/28/10/313",
    journal = "J. Phys. G",
    volume = "28",
    pages = "2693",
    year = "2002"
}

@article{Muller:1993ig,
    author = "Muller, U. and Engelfried, J. and Gerassimov, S. G. and Martens, K. and Michaels, R. and Siebert, H. W. and Walder, G.",
    editor = "Nappi, E. and Ypsilantis, T.",
    title = "{Particle identification with the RICH detector in experiment WA89 at CERN}",
    reportNumber = "CERN-PPE-93-109",
    doi = "10.1016/0168-9002(94)90565-7",
    journal = "Nucl. Instrum. Meth. A",
    volume = "343",
    pages = "279",
    year = "1994"
}

@article{cortina2023performance,
  title="{Performance of the NA62 trigger system}",
  author={Cortina Gil, Eduardo and others},
  journal={JHEP},
  volume={03},
  pages={122},
  year={2023},
  doi = {10.1007/JHEP03(2023)122},
  publisher={Springer}
}

@article{ahmad2002direct,
  title={Direct evidence for neutrino flavor transformation from neutral-current interactions in the Sudbury Neutrino Observatory},
  author={Ahmad, Q Retal and others},
  journal={Phys. Rev. Lett.},
  volume={89},
  number={1},
  pages={011301},
  year={2002},
  doi = {10.1103/PhysRevLett.89.011301},
  publisher={APS}
}

@article{hirata1988experimental,
  title={Experimental study of the atmospheric neutrino flux},
  author={Hirata, KS  and others},
  journal={Phys. Lett. B},
  volume={205},
  pages={416},
  year={1988},
  doi = {10.1016/0370-2693(88)91690-5},
  publisher={Elsevier}
}

@article{asaka2005numsm,
  title="{The $\nu$MSM, dark matter and baryon asymmetry of the universe}",
  author={Asaka, Takehiko and Shaposhnikov, Mikhail},
  journal={Phys. Lett. B},
  volume={620},
  pages={17},
  year={2005},
  doi = {10.1016/j.physletb.2005.06.020},
  publisher={Elsevier}
}

@article{asaka-2,
  title="{The $\nu$MSM, dark matter and neutrino masses}",
  author={Asaka, Takehiko and Blanchet, Steve and Shaposhnikov, Mikhail},
  journal={Phys. Lett. B},
  volume={631},
  pages={151},
  year={2005},
  doi = {10.1016/j.physletb.2005.09.070},
  publisher={Elsevier}
}

@article{MINKOWSKI1977421,
title = {$\mu\rightarrow e\gamma$ at a rate of one out of $10^{9}$ muon decays?},
journal = {Phys. Lett. B},
volume = {67},
pages = {421},
year = {1977},
issn = {0370-2693},
doi = {10.1016/0370-2693(77)90435-X},
author = {Peter Minkowski}
}

@article{shrock1980new,
  title={New tests for and bounds on neutrino masses and lepton mixing},
  author={Shrock, Robert E},
  journal={Phys. Lett. B},
  volume={96},
  pages={159},
  year={1980},
  doi = "10.1016/0370-2693(80)90235-X",
  publisher={Elsevier}
}

@article{NA62:2023mud,
    author = "Bethani, Agni and others",
    collaboration = "NA62",
    title = "{Development of a new CEDAR for kaon identification at the NA62 experiment at CERN}",
    eprint = "2312.17188",
    archivePrefix = "arXiv",
    primaryClass = "hep-ex",
    reportNumber = "CERN-EP-2023-302",
    doi = "10.1088/1748-0221/19/05/P05005",
    journal = "JINST",
    volume = "19",
    pages = "P05005",
    year = "2024"
}

\newpage

\newcommand{\orcimg}{\raisebox{-0.3\height}{\includegraphics[height=\fontcharht\font`A]{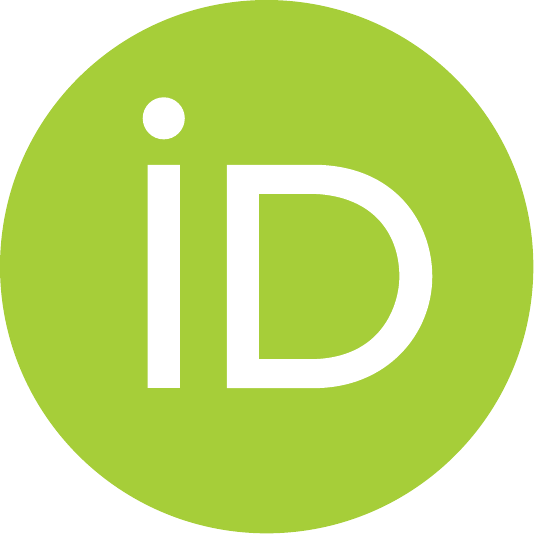}}}
\newcommand{\orcid}[1]{\href{https://orcid.org/#1}{\orcimg}}

\centerline {\bf The NA62 Collaboration}
\vspace{1.5cm}
%
%

\begin{raggedright}
\noindent
{\bf Universit\'e Catholique de Louvain, Louvain-La-Neuve, Belgium}\\
 B.~Bloch-Devaux$\,${\footnotemark[1]}\orcid{0000-0002-2463-1232},
 E.~Cortina Gil\orcid{0000-0001-9627-699X},
 N.~Lurkin\orcid{0000-0002-9440-5927},
 E.~Minucci$\,${\footnotemark[2]}\orcid{0000-0002-3972-6824},
 S.~Padolski\orcid{0000-0002-6795-7670},
 P.~Petrov
\vspace{0.5cm}

{\bf TRIUMF, Vancouver, British Columbia, Canada}\\
 T.~Numao\orcid{0000-0001-5232-6190},
 Y.~Petrov\orcid{0000-0003-2643-8740},
 V.~Shang\orcid{0000-0002-1436-6092},
 B.~Velghe\orcid{0000-0002-0797-8381},
 V. W. S.~Wong\orcid{0000-0001-5975-8164}
% ...
\vspace{0.5cm}

{\bf Instituto de F\'isica, Universidad Aut\'onoma de San Luis Potos\'i, San Luis Potos\'i, Mexico}\\
 A.~Briano Olvera$\,$\renewcommand{\thefootnote}{\fnsymbol{footnote}}\footnotemark[1]\renewcommand{\thefootnote}{\arabic{footnote}}\orcid{0000-0001-6121-3905},
 J.~Engelfried$\,$\renewcommand{\thefootnote}{\fnsymbol{footnote}}\footnotemark[1]\renewcommand{\thefootnote}{\arabic{footnote}}\orcid{0000-0001-5478-0602},
 N.~Estrada-Tristan$\,${\footnotemark[21]}\orcid{0000-0003-2977-9380},
 R.~Piandani\orcid{0000-0003-2226-8924},
 M.~A.~Reyes~Santos$\,${\footnotemark[21]}\orcid{0000-0003-1347-2579},
 K.~A.~Rodriguez~Rivera\orcid{0000-0001-5723-9176}
\vspace{0.5cm}

\newpage
{\bf Horia Hulubei National Institute for R\&D in Physics and Nuclear Engineering, Bucharest-Magurele, Romania}\\
 P.~Boboc\orcid{0000-0001-5532-4887},
 A. M.~Bragadireanu,
 S. A.~Ghinescu\orcid{0000-0003-3716-9857},
 O. E.~Hutanu
\vspace{0.5cm}

{\bf Faculty of Mathematics, Physics and Informatics, Comenius University, Bratislava, Slovakia}\\
 T.~Blazek\orcid{0000-0002-2645-0283},
 V.~Cerny\orcid{0000-0003-1998-3441},
 T.~Velas\orcid{0009-0004-0061-1968},
 R.~Volpe$\,${\footnotemark[22]}\orcid{0000-0003-1782-2978}
\vspace{0.5cm}

{\bf CERN, European Organization for Nuclear Research, Geneva, Switzerland}\\
 J.~Bernhard\orcid{0000-0001-9256-971X},
 M.~Boretto\orcid{0000-0001-5012-4480},
 F.~Brizioli$\,${\footnotemark[22]}\orcid{0000-0002-2047-441X},
 A.~Ceccucci\orcid{0000-0002-9506-866X},
 H.~Danielsson\orcid{0000-0002-1016-5576}, 
 N.~De Simone$\,${\footnotemark[23]},
 F.~Duval,
 L.~Federici$\,${\footnotemark[24]}\orcid{0000-0002-3401-9522},
 E.~Gamberini\orcid{0000-0002-6040-4985},
 R.~Guida, 
 F.~Hahn$\,$\renewcommand{\thefootnote}{\fnsymbol{footnote}}\footnotemark[2]\renewcommand{\thefootnote}{\arabic{footnote}},
 E. B.~Holzer\orcid{0000-0003-2622-6844},
 B.~Jenninger,
 Z.~Kucerova\orcid{0000-0001-8906-3902},
 P.~Laycock$\,${\footnotemark[25]}\orcid{0000-0002-8572-5339}, 
 G.~Lehmann Miotto\orcid{0000-0001-9045-7853},
 P.~Lichard\orcid{0000-0003-2223-9373},
 A.~Mapelli\orcid{0000-0002-4128-1019},
 M.~Noy,
 V.~Palladino\orcid{0000-0002-9786-9620}, 
 V.~Ryjov,
 S.~Venditti,
 M.~Zamkovsky\orcid{0000-0002-5067-4789}
\vspace{0.5cm}

{\bf Ecole Polytechnique F\'ed\'erale Lausanne, Lausanne, Switzerland}\\
 X.~Chang\orcid{0000-0002-8792-928X},
 A.~Kleimenova\orcid{0000-0002-9129-4985},
 R.~Marchevski\orcid{0000-0003-3410-0918}
\vspace{0.5cm}

{\bf School of Physics and Astronomy, University of Birmingham, Birmingham, United Kingdom}\\
 T.~Bache\orcid{0000-0003-4520-830X},
 M. B.~Brunetti$\,${\footnotemark[26]}\orcid{0000-0003-1639-3577},
 V.~Fascianelli$\,${\footnotemark[27]},
 J. R.~Fry\orcid{0000-0002-3680-361X},
 F.~Gonnella\orcid{0000-0003-0885-1654}, 
 E.~Goudzovski\orcid{0000-0001-9398-4237},
 J.~Henshaw\orcid{0000-0001-7059-421X},
 L.~Iacobuzio,
 C.~Kenworthy\orcid{0009-0002-8815-0048},
 C.~Lazzeroni\orcid{0000-0003-4074-4787}, 
 F.~Newson,
 C.~Parkinson\orcid{0000-0003-0344-7361},
 A.~Romano\orcid{0000-0003-1779-9122},
 C.~Sam\orcid{0009-0005-3802-5777},
 J.~Sanders\orcid{0000-0003-1014-094X}, 
 A.~Sergi$\,${\footnotemark[28]}\orcid{0000-0001-9495-6115},
 A.~Sturgess\orcid{0000-0002-8104-5571},
 A.~Tomczak\orcid{0000-0001-5635-3567}
\vspace{0.5cm}

{\bf School of Physics, University of Bristol, Bristol, United Kingdom}\\
 H.~Heath\orcid{0000-0001-6576-9740},
 R.~Page,
 S.~Trilov\orcid{0000-0003-0267-6402}
\vspace{0.5cm}

{\bf School of Physics and Astronomy, University of Glasgow, Glasgow, United Kingdom}\\
 B.~Angelucci,
 D.~Britton\orcid{0000-0001-9998-4342},
 C.~Graham\orcid{0000-0001-9121-460X},
 A.~Norton\orcid{0000-0001-5959-5879},
 D.~Protopopescu\orcid{0000-0002-8047-6513}
\vspace{0.5cm}

{\bf Physics Department, University of Lancaster, Lancaster, United Kingdom}\\
 J.~Carmignani$\,${\footnotemark[29]}\orcid{0000-0002-1705-1061},
 J. B.~Dainton,
 L.~Gatignon\orcid{0000-0001-6439-2945},
 R. W. L.~Jones\orcid{0000-0002-6427-3513},
 K.~Massri\orcid{0000-0001-7533-6295},
 A.~Shaikhiev\orcid{0000-0003-2921-8743}
\vspace{0.5cm}

{\bf School of Physical Sciences, University of Liverpool, Liverpool, United Kingdom}\\
 L.~Fulton,
 D.~Hutchcroft\orcid{0000-0002-4174-6509},
 E.~Maurice$\,${\footnotemark[30]}\orcid{0000-0002-7366-4364},
 B.~Wrona\orcid{0000-0002-1555-0262}
\vspace{0.5cm}

{\bf Physics and Astronomy Department, George Mason University, Fairfax, Virginia, USA}\\
 A.~Conovaloff,
 P.~Cooper,
 D.~Coward$\,${\footnotemark[31]}\orcid{0000-0001-7588-1779},
 P.~Rubin\orcid{0000-0001-6678-4985}
\vspace{0.5cm}

{\bf Authors affiliated with an international laboratory covered by a cooperation agreement with CERN}\\
 A.~Baeva,
 D.~Baigarashev$\,${\footnotemark[32]}\orcid{0000-0001-6101-317X},
 V.~Bautin\orcid{0000-0002-5283-6059},
 D.~Emelyanov,
 T.~Enik\orcid{0000-0002-2761-9730}, 
 K.~Gorshanov\orcid{0000-0001-7912-5962},
 V.~Kekelidze\orcid{0000-0001-8122-5065},
 D.~Kereibay,
 A.~Korotkova,
 L.~Litov$\,${\footnotemark[12]}\orcid{0000-0002-8511-6883}, 
 D.~Madigozhin\orcid{0000-0001-8524-3455},
 M.~Misheva$\,${\footnotemark[33]},
 N.~Molokanova,
 S.~Movchan,
 A.~Okhotnikov\orcid{0000-0003-1404-3522}, 
 I.~Polenkevich,
 Yu.~Potrebenikov\orcid{0000-0003-1437-4129},
 A.~Sadovskiy\orcid{0000-0002-4448-6845},
 K.~Salamatin\orcid{0000-0001-6287-8685},
 S.~Shkarovskiy, 
 A.~Zinchenko$\,$\renewcommand{\thefootnote}{\fnsymbol{footnote}}\footnotemark[2]\renewcommand{\thefootnote}{\arabic{footnote}}
\vspace{0.5cm}

\newpage
{\bf Authors affiliated with an Institute formerly covered by a cooperation agreement with CERN}\\
 S.~Fedotov,
 E.~Gushchin\orcid{0000-0001-8857-1665},
 A.~Khotyantsev,
 Y.~Kudenko\orcid{0000-0003-3204-9426},
 V.~Kurochka, 
 V.~Kurshetsov\orcid{0000-0003-0174-7336},
 M.~Medvedeva,
 A.~Mefodev,
 V.~Obraztsov\orcid{0000-0002-0994-3641},
 A.~Ostankov$\,$\renewcommand{\thefootnote}{\fnsymbol{footnote}}\footnotemark[2]\renewcommand{\thefootnote}{\arabic{footnote}}, 
 V.~Semenov$\,$\renewcommand{\thefootnote}{\fnsymbol{footnote}}\footnotemark[2]\renewcommand{\thefootnote}{\arabic{footnote}},
 V.~Sugonyaev\orcid{0000-0003-4449-9993},
 O.~Yushchenko\orcid{0000-0003-4236-5115}
\vspace{0.5cm}

\end{raggedright}

%
%

\setcounter{footnote}{0}
\newlength{\basefootnotesep}
\setlength{\basefootnotesep}{\footnotesep}

\renewcommand{\thefootnote}{\fnsymbol{footnote}}
\noindent
$^{\footnotemark[1]}${Corresponding authors: A.~Briano Olvera, J.~Engelfried, \\
email: alejandro.briano.olvera@cern.ch, jurgen.engelfried@cern.ch,}\\
$^{\footnotemark[2]}${Deceased}\\
\renewcommand{\thefootnote}{\arabic{footnote}}
$^{1}${Also at School of Physics and Astronomy, University of Birmingham, Birmingham, B15 2TT, UK} \\
$^{2}${Present address: INFN, Laboratori Nazionali di Frascati, I-00044 Frascati, Italy} \\
$^{3}${Also at TRIUMF, Vancouver, British Columbia, V6T 2A3, Canada} \\
$^{4}${Also at Universit\'e de Toulon, Aix Marseille University, CNRS, IM2NP, F-83957, La Garde, France} \\
$^{5}${Also at Department of Physics, Technical University of Munich, M\"unchen, D-80333, Germany} \\
$^{6}${Present address: Institut f\"ur Kernphysik and Helmholtz Institute Mainz, Universit\"at Mainz, Mainz, D-55099, Germany} \\
$^{7}${Also at CERN, European Organization for Nuclear Research, CH-1211 Geneva 23, Switzerland} \\
$^{8}${Present address: Universit\"at W\"urzburg, D-97070 W\"urzburg, Germany} \\
$^{9}${Present address: European XFEL GmbH, D-22869 Schenefeld, Germany} \\
$^{10}${Also at Dipartimento di Scienze Fisiche, Informatiche e Matematiche, Universit\`a di Modena e Reggio Emilia, I-41125 Modena, Italy} \\
$^{11}${Present address: Max-Planck-Institut f\"ur Physik (Werner-Heisenberg-Institut), Garching, D-85748, Germany} \\
$^{12}${Also at Faculty of Physics, University of Sofia, BG-1164 Sofia, Bulgaria} \\
$^{13}${Also at INFN, Sezione di Roma I e Dipartimento di Fisica, Sapienza Universit\`a di Roma, I-00185 Roma, Italy} \\
$^{14}${Present address: INFN, Sezione di Napoli e Scuola Superiore Meridionale, I-80138 Napoli, Italy} \\
$^{15}${Present address: CERN, European Organization for Nuclear Research, CH-1211 Geneva 23, Switzerland} \\
$^{16}${Also at Department of Industrial Engineering, University of Roma Tor Vergata, I-00173 Roma, Italy} \\
$^{17}${Also at Department of Electronic Engineering, University of Roma Tor Vergata, I-00173 Roma, Italy} \\
$^{18}${Also at Gran Sasso Science Institute, I-67100 L'Aquila,  Italy} \\
$^{19}${Also at Universit\`a degli Studi del Piemonte Orientale, I-13100 Vercelli, Italy} \\
$^{20}${Also at al-Farabi Kazakh National University, 050040 Almaty, Kazakhstan} \\
$^{21}${Also at Universidad de Guanajuato, 36000 Guanajuato, Mexico} \\
$^{22}${Present address: INFN, Sezione di Perugia, I-06100 Perugia, Italy} \\
$^{23}${Present address: DESY, D-15738 Zeuthen, Germany} \\
$^{24}${Present address: IPHC, CNRS/IN2P3, Strasbourg University, F-67037 Strasbourg, France} \\
$^{25}${Present address: Brookhaven National Laboratory, Upton, NY 11973, USA} \\
$^{26}${Present address: Kansas University, Lawrence, KS 66045-7582, USA} \\
$^{27}${Present address: Center for theoretical neuroscience, Columbia University, New York, NY 10027, USA} \\
$^{28}${Present address: INFN, Sezione di Genova e Dipartimento di Fisica dell'Universit\`a, I-16146 Genova, Italy} \\
$^{29}${Present address: School of Physical Sciences, University of Liverpool, Liverpool, L69 7ZE, UK} \\
$^{30}${Present address: Laboratoire Leprince Ringuet, F-91120 Palaiseau, France} \\
$^{31}${Also at SLAC National Accelerator Laboratory, Stanford University, Menlo Park, CA 94025, USA} \\
$^{32}${Also at L. N. Gumilyov Eurasian National University, 010000 Nur-Sultan, Kazakhstan} \\
$^{33}${Present address: Institute of Nuclear Research and Nuclear Energy of Bulgarian Academy of Science (INRNE-BAS), BG-1784 Sofia, Bulgaria} \\

\end{document}